\documentclass[prd,twocolumn,showpacs]{revtex4}
\usepackage{amsmath,bm,graphicx,slashed}

\begin{document}

\title{2-soft-gluon exchange and factorization breaking}
\author{John Collins}
\email{collins@phys.psu.edu}
\affiliation{
    Physics Department,
    Penn State University,
    104 Davey Laboratory,
    University Park PA 16802,
    U.S.A.
}
\date{31 August 2007}

\begin{abstract}
  A previous counterexample to disprove $k_T$-factorization for
  $H_1+H_2 \to H_3+H_4+X$ is extended calculationally to one higher
  order in gluon exchange.  The result is that, by explicit
  calculation, standard $k_T$-factorization fails for the unpolarized
  cross-section for the production of hadrons of high transverse
  momentum in hadron-hadron collisions.
\end{abstract}


\maketitle

\section{Introduction}

Hard-scattering factorization --- both conventional collinear
factorization and $k_T$-factorization --- is of great phenomenological
importance in QCD phenomenology.  It is therefore very important that
it has been found, by the Amsterdam group
\cite{bomhof_04,Bacchetta:2005rm,Bomhof:2006dp,Pijlman:2006tq}, that
parton densities appear to be non-universal and process-dependent for
the production of high transverse momentum hadrons in hadron-hadron
collisions: $H_1+H_2 \to H_3+H_4+X$.  This is for the case that the
detected hadrons are close to back-to-back azimuthally, so that
$k_T$-factorization, and hence transverse-momentum-dependent (TMD)
parton densities and fragmentation functions, are to be used.  

The changes in the parton densities involve unusual paths for the
Wilson lines in their operator definitions.  Although the use of these
paths is quite natural, it was not completely obvious that, for
example, standard factorization could not also be valid, with some
nontrivial transformation relating the different kinds of TMD
functions.  So recently, Collins and Qiu \cite{Collins:2007nk}
constructed a counterexample simple enough to show that this could not
be the case.  The simplicity of the counterexample was partly due to
its application to a transverse single-spin asymmetry (SSA).

Meanwhile another approach, by Qiu, Vogelsang and Yuan, culminating in
Refs.\ \cite{Qiu:2007ar,Qiu:2007ey}, led to an apparently opposite
result.  This was that standard factorization could be valid for the
SSA provided that the hard scattering factor is redefined.  The
contradiction is particularly striking because the model formulated in
\cite{Collins:2007nk} as a counterexample to factorization is one to
which the arguments of \cite{Qiu:2007ar,Qiu:2007ey} in favor of
factorization also clearly apply.

Therefore it is the purpose of this paper to lay to rest any doubts
about nonfactorization by extending the counterexample of
\cite{Collins:2007nk} to one higher order of perturbation theory.  The
methods of \cite{Collins:2007nk} enable this calculation to be done
quite simply.

First, certain terminological issues need to be addressed.  In
\cite{Collins:2007nk}, as in the present paper, ``factorization''
means ``standard factorization''.  That is, the nonperturbative parton
densities and fragmentation functions for the $H_1+H_2 \to H_3+H_4+X$
process either are those extracted from $e^+e^-$ annihilation and from
reactions in deep-inelastic scattering (DIS), or are related to them
by purely perturbatively-based calculations.  This is important for
phenomenology, since perturbative calculations of hard-scattering
coefficients then give predictions from first principles, to a useful
degree of accuracy.  In contrast, Refs.\
\cite{bomhof_04,Bacchetta:2005rm,Bomhof:2006dp,Pijlman:2006tq} find a
more general factorization with a greater variety of
reaction-dependent nonperturbative functions.  Such a factorization
is not standard factorization in the sense just defined.

Notice that the notorious reversal of sign of the Sivers function
between DIS and Drell-Yan (DY) processes \cite{collins_02} is already
pushing the limits of what can be accommodated under this definition
of standard factorization: the actual operator definitions of the
parton densities for the two processes are definitively different, and
are only related numerically because of the time-reversal symmetry of
QCD.

The key technical issue is that, at the level of \emph{individual}
Feynman graphs, there are extra leading-power exchanges of gluons
between the subgraphs that correspond to the different factors in a
statement of factorization.  Factorization only holds after
application of appropriate approximations followed by application of
Ward identities to extract the extra gluons in particular kinematic
regions from their attachments to the interior of subgraphs for other
kinematic regions.  Thus only after a sum over graphs can one obtain
factorization, provided that the operator definitions of the parton
densities and fragmentation functions are equipped with suitably
compatible Wilson lines.  A Wilson-line operator is the exponential of
its one-gluon term; thus the use of Wilson lines implies certain
relations between the values of Feynman graphs with different numbers
of exchanged gluons.

The necessary approximations are only valid after certain contour
deformations are applied to the momentum integrals.  The results of
\cite{bomhof_04,Bacchetta:2005rm,Bomhof:2006dp,Pijlman:2006tq} are
essentially that the pattern of initial- and final-state parton lines
in the process $H_1+H_2 \to H_3+H_4+X$ is appropriate to contour
deformations different from those appropriate to standard
factorization.  In \cite{Collins:2007nk}, particular graphs for the
single-spin asymmetry (SSA) were calculated, and led to a result that
is inconsistent with standard factorization.  The factor for one
parton density depends on the color charge of the other parton(s)
participating in the process.  Any factorization must be of the more
general form found in
\cite{bomhof_04,Bacchetta:2005rm,Bomhof:2006dp,Pijlman:2006tq}, where
the nonperturbative physics must be contained in extra functions
defined with more complicated Wilson lines.

I will first explain, in Sec.\ \ref{sec:QVY}, how to reconcile this
with the apparently opposing conclusions derived from compatible
Feynman graph calculations in \cite{Qiu:2007ar,Qiu:2007ey}.  That this
discussion is rather abstract, motivates the calculations I present in
Sec.\ \ref{sec:2ge}.  There I calculate the effects of the exchange of
an extra gluon as compared with the calculations in
\cite{Collins:2007nk}.  I will find an explicit failure of the Wilson
line exponentiation for the unpolarized cross section.  I will also
point out that a corresponding result for the SSA will need yet an
extra exchanged gluon.

\section{Comparison with \cite{Qiu:2007ar,Qiu:2007ey}}
\label{sec:QVY}

In this section, we examine the relation between the result of
nonfactorization in \cite{Collins:2007nk} and the argument compatible
with factorization in \cite{Qiu:2007ar,Qiu:2007ey}, which appears to
encompass the model calculations in \cite{Collins:2007nk}.

\subsection{Direct comparison}
\label{sec:comparison}

\begin{figure}
  \centering
    \includegraphics[scale=0.4]{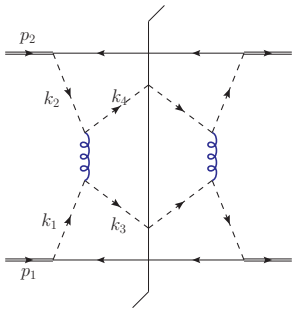}
    \caption{(Color online.)
      Lowest order graph in the model for hadroproduction of
      hadrons of high transverse momentum.  The initial state
      particles are color-singlet Dirac particles.  The spectator
      lines are for Dirac ``quark'' fields of charges $g_1$ and $g_2$,
      and the active partons are for scalar ``diquark'' fields.  The
      exchanged gluon line is thickened to denote the hard scattering.
    }
  \label{fig:HHHH0}
\end{figure}

\begin{figure}
  \centering
  \includegraphics[scale=0.4]{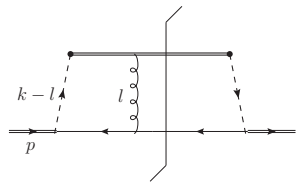}
  \caption{Virtual one-gluon-exchange correction to parton
    density. The upper double line is the Wilson line, and the graph
    shown, together with its Hermitian conjugate gives the first
    contribution to the Sivers function.}
  \label{fig:pdf1}
\end{figure}

\begin{figure*}
  \centering
  \begin{tabular}{c@{\hspace*{5mm}}c@{\hspace*{5mm}}c}
    \includegraphics[scale=0.4]{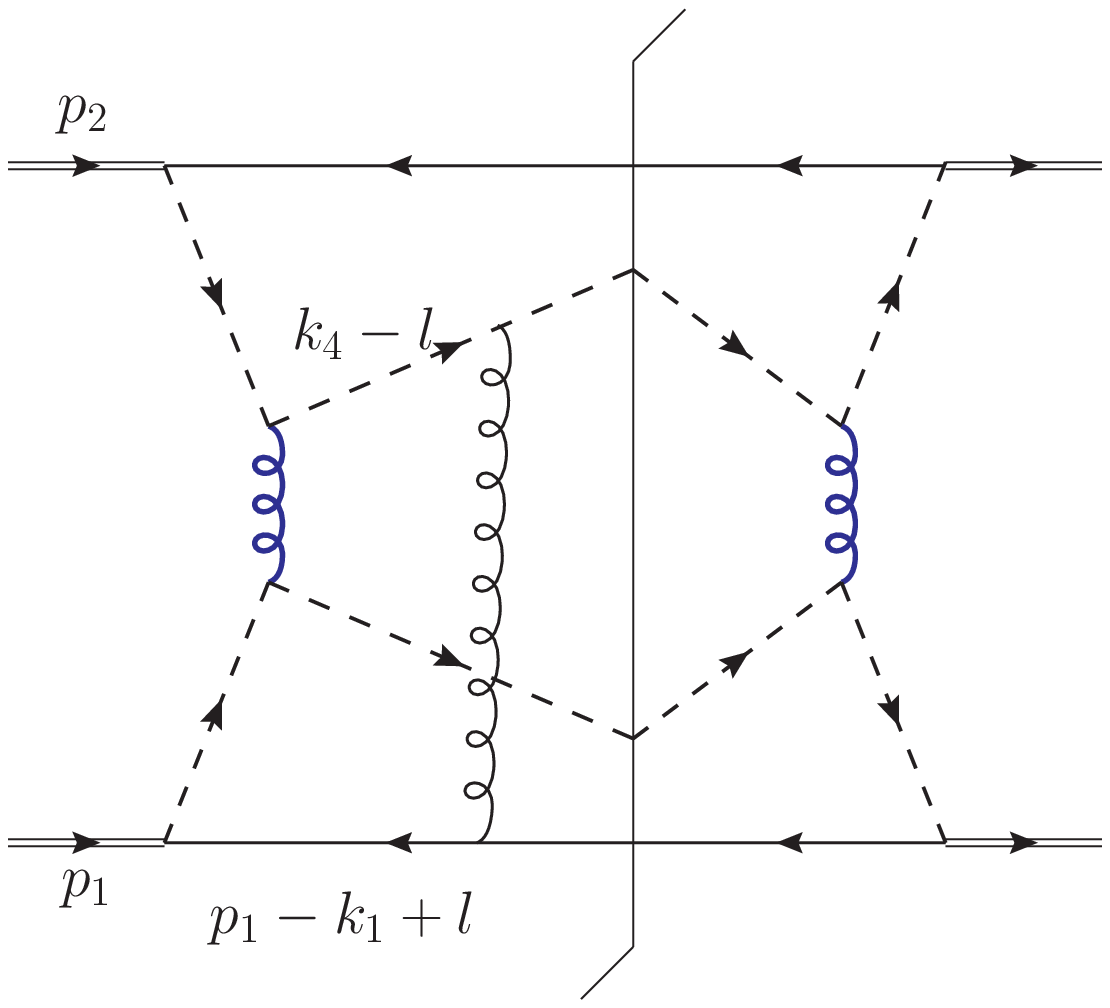}
  &
    \includegraphics[scale=0.4]{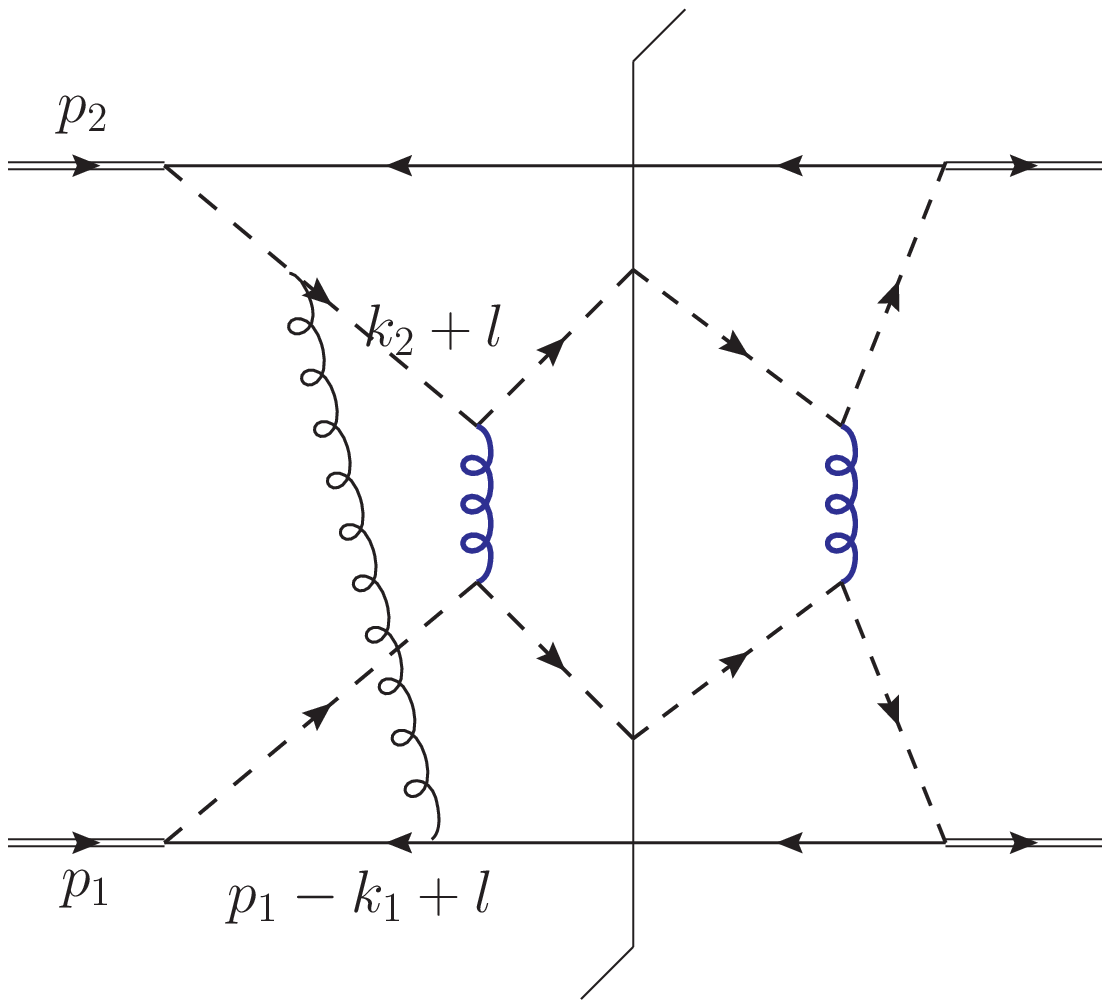}
  &
    \includegraphics[scale=0.4]{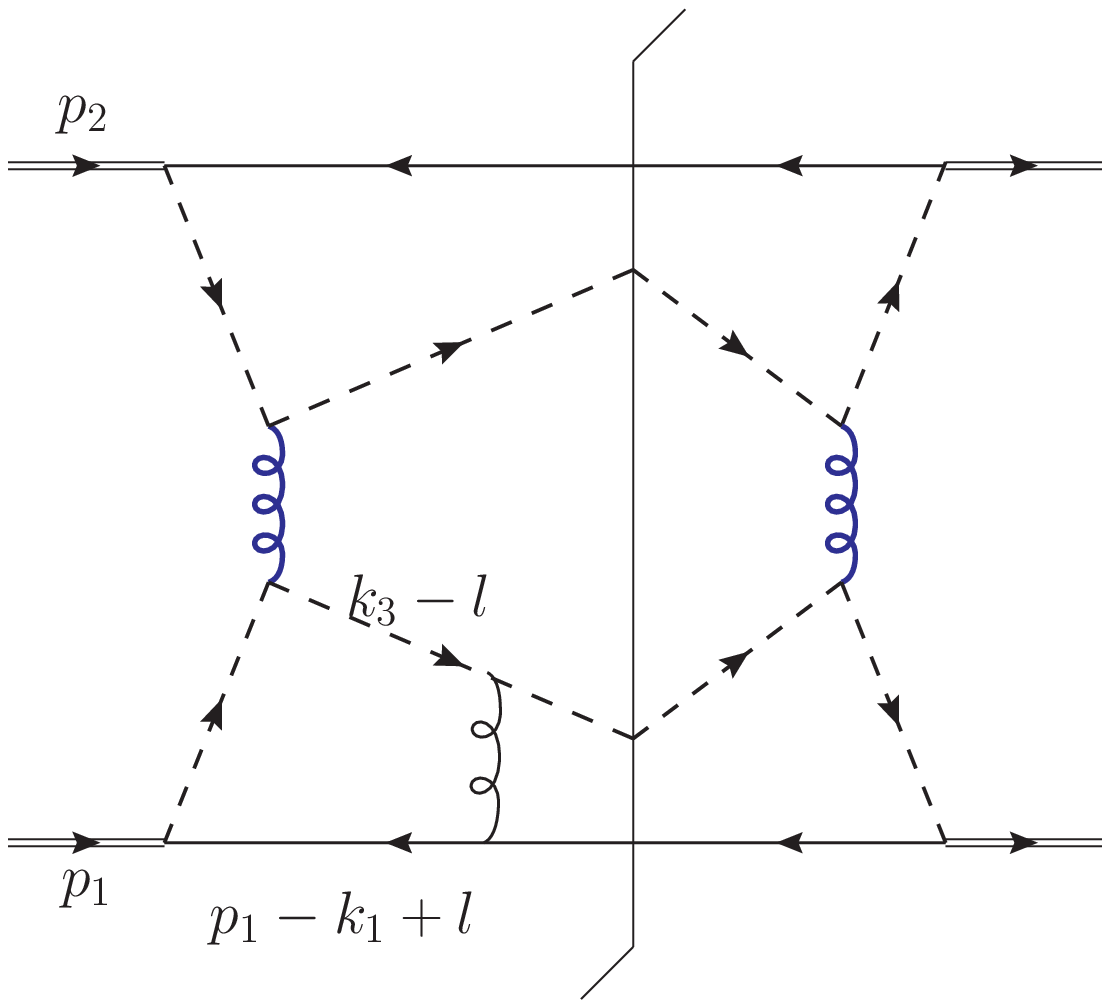}
  \\
    (a) & (b) & (c)
  \end{tabular}
  \caption{(Color online.)
    Exchange of one extra gluon. Only graphs are shown that are
    relevant to the connection of the lower spectator line to the
    Wilson line in the associated parton density.  Hermitian
    conjugates of these graphs also contribute, to give a total of 6
    graphs.  }
  \label{fig:HHHH1}
\end{figure*}

Recall that in the model of \cite{Collins:2007nk}, the lowest-order
graph for the process $H_1+H_2 \to H_3+H_4+X$ is Fig.\ \ref{fig:HHHH0},
with a single hard gluon exchange.  To obtain the putative Wilson-line
contribution to the parton density in the lower hadron with one
virtual gluon connecting the Wilson line to the spectator line, Fig.\
\ref{fig:pdf1}, we need to sum the graphs of Fig.\ \ref{fig:HHHH1}.
For the \emph{real} part of the amplitude, the result does indeed
correspond to standard factorization; i.e., the summed one-gluon
correction in the cross section corresponds to the one-gluon
correction to the parton density Fig.\ \ref{fig:pdf1}.

However for the \emph{imaginary} part of the amplitude, there is a
mismatch by a factor that depends on the colors of the active partons
for the hard scattering.  Thus there is a failure of the steps used in
proving standard factorization: a combination of contour deformation,
approximation, and a Ward identity.  The imaginary part of the
amplitude gives the lowest order SSA, so the effect at this order is
manifested in the SSA but not in the unpolarized cross section.
\emph{If one restricts attention to this order in the number of
  exchanged gluons}, then, as observed in
\cite{Qiu:2007ar,Qiu:2007ey}, one obtains factorization for the SSA
simply by multiplying the hard scattering by the appropriate color
factor.  The calculated hard scattering coefficient is different
between the SSA and the unpolarized cross section.

However if standard factorization were true, graphs with more gluon
exchanges would have to organize themselves into an exponentiated
Wilson line operator with the color charge appropriate to the struck
parton, and with the calculated color factor from Fig.\
\ref{fig:HHHH1} being the same for all the higher-order terms.

The only known argument for this is the same kind of Ward identity
argument that is used in standard factorization proofs
\cite{bodwin_85,collins_85_88}.  The result of Collins and Qiu
\cite{Collins:2007nk} is that the Ward identity argument fails.  In
contrast, Qiu, Vogelsang and Yuan \cite{Qiu:2007ar,Qiu:2007ey}
explicitly leave the effect of extra gluon exchange to future work.

\subsection{Importance of multiple gluon exchange}
\label{sec:meta.proof}

It is fairly easy to miss the central logical point of
\cite{Collins:2007nk}.  For example, Ratcliffe and Teryaev
\cite{Ratcliffe:2007ye} state ``The main point of the argument in
\cite{Collins:2007nk} is the proportionality of the contribution of
the Sivers function to the electric charge of the quark from the other
(unpolarised) hadron.''  That much was previously quite well-known
from other work, e.g.,
\cite{bomhof_04,Bacchetta:2005rm,Bomhof:2006dp,Pijlman:2006tq,%
  Qiu:2007ar,Qiu:2007ey}.  The contribution of \cite{Collins:2007nk}
was to show in as elementary and transparent a fashion as the authors
could manage that this fact must be interpreted as a breakdown of
standard factorization, rather than as giving a changed normalization
for the hard-scattering coefficient for the SSA.  Moreover the failure
of factorization is not just for the SSA, but also for the unpolarized
cross section.

In view of the importance of such issues to this paper, I now
re-emphasize them here.

The proof in \cite{Collins:2007nk} is really a meta-proof, a proof
about proofs.  To get factorization, one must extract, into Wilson
lines, exchanges of arbitrarily many gluons between collinear and hard
subgraphs.  A direct calculation is evidently impractical, and any
proof relies on more general methods, i.e., Ward identities.  The
actual calculation in \cite{Collins:2007nk} shows that the requisite
Ward identity fails quantitatively.  Therefore even though a
computation at low order is compatible with factorization, the
arguments needed to extend factorization to all orders do not work.

Methods for deriving factorization are general across field theories.
If they work in a complicated theory like QCD, they also work in a
simple model theory.  Conversely and most importantly, if methods fail
in the model, they will also fail in QCD.  This again is a meta-proof. 

Further issues are that there are many graphs of the order considered,
many more beyond those actually considered, Fig.\ \ref{fig:HHHH1}, and
that an exact calculation of the graphs is hard.  So in arguments
claiming to demonstrate nonfactorization from an examination of a
limited set of graphs, one has to be concerned whether some other
graphs matter, or whether an inappropriate approximation was used, or
whether the graphs can be interpreted differently in terms of
factorization.

Therefore in \cite{Collins:2007nk}, the model and process were
carefully chosen so that only the graphs in Fig.\ \ref{fig:HHHH1} were
relevant, with this being particularly clear for the SSA.

\section{Exchange of extra gluons}
\label{sec:2ge}

The model has an Abelian massive gluon, and a reaction is examined in
which beam particles are color-neutral, and the partons in the lower
and upper hadrons (e.g., in Fig.\ \ref{fig:HHHH0}) have charges $g_1$
and $g_2$.

\begin{figure*}
  \centering
  \begin{tabular}{c@{\hspace*{5mm}}c@{\hspace*{5mm}}c}
    \includegraphics[scale=0.4]{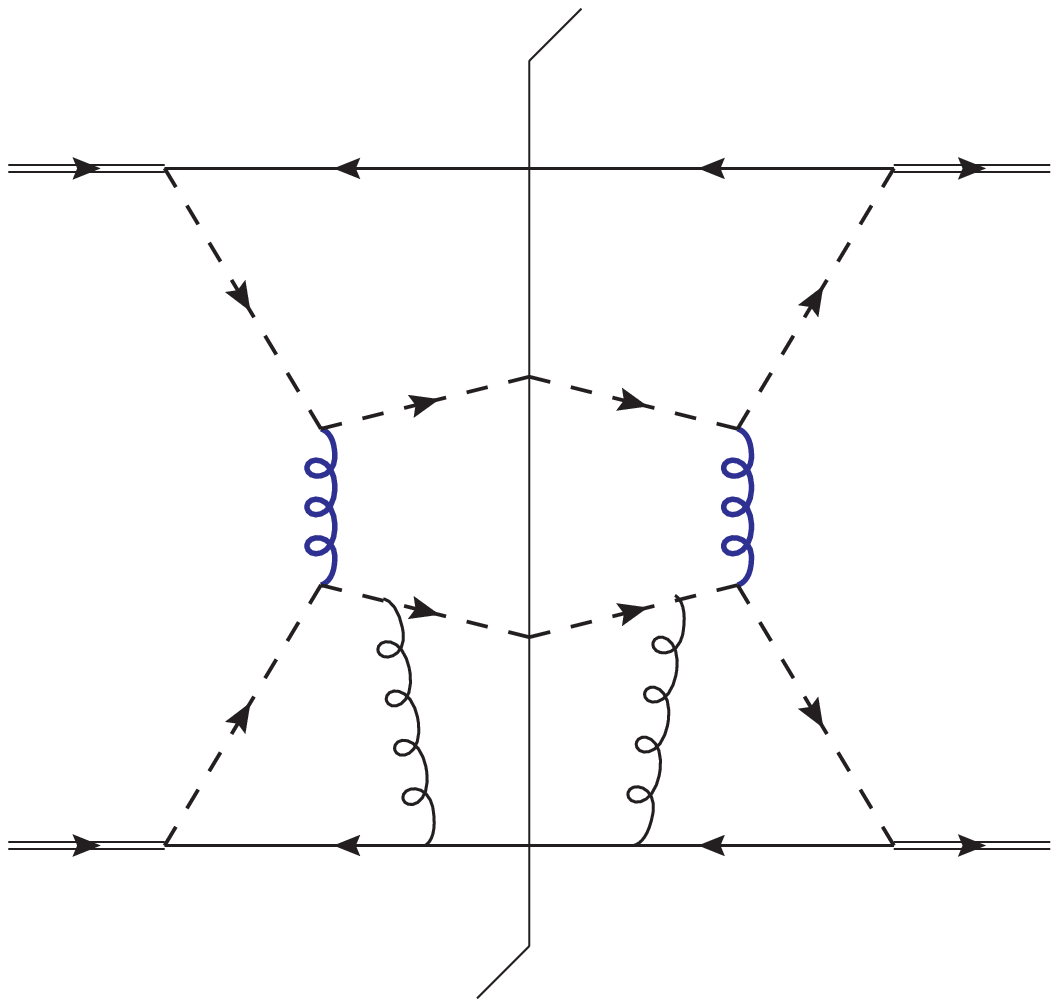}
  &
    \includegraphics[scale=0.4]{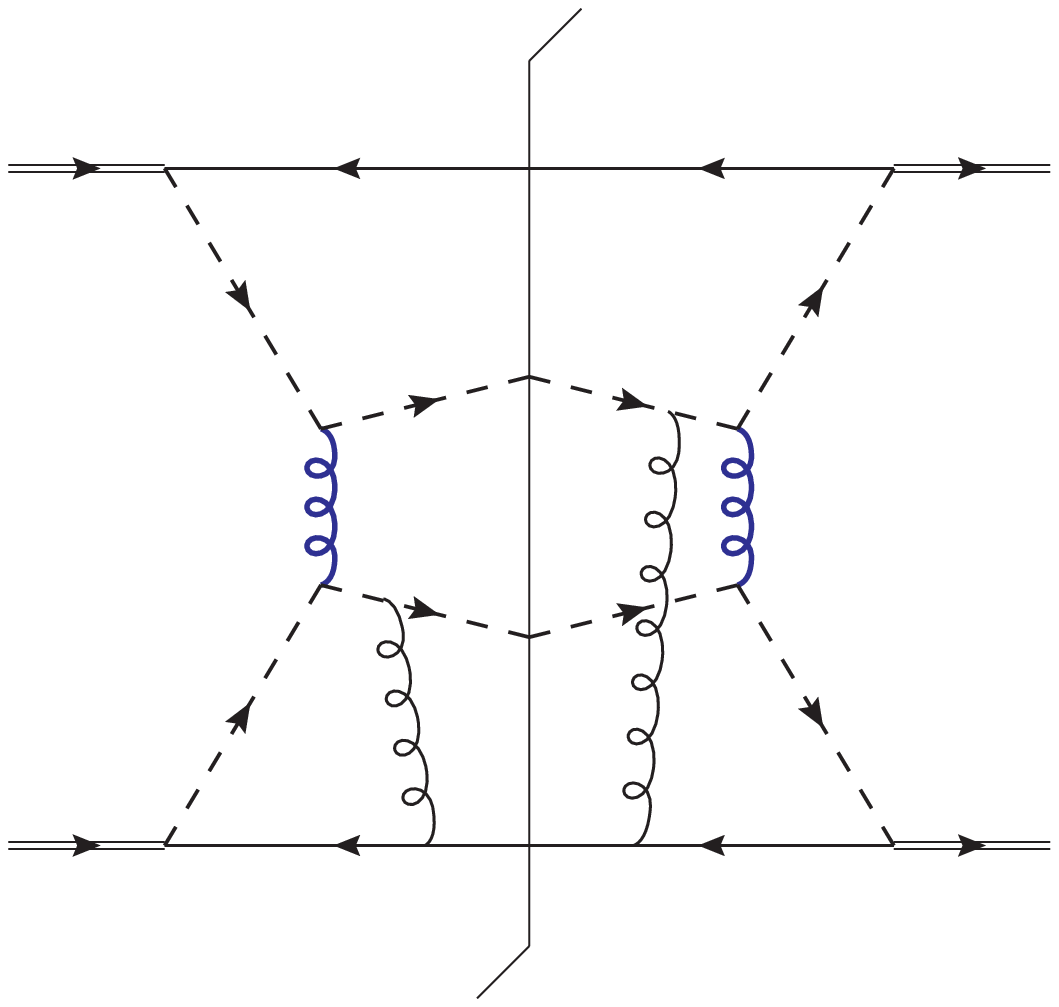}
  &
    \includegraphics[scale=0.4]{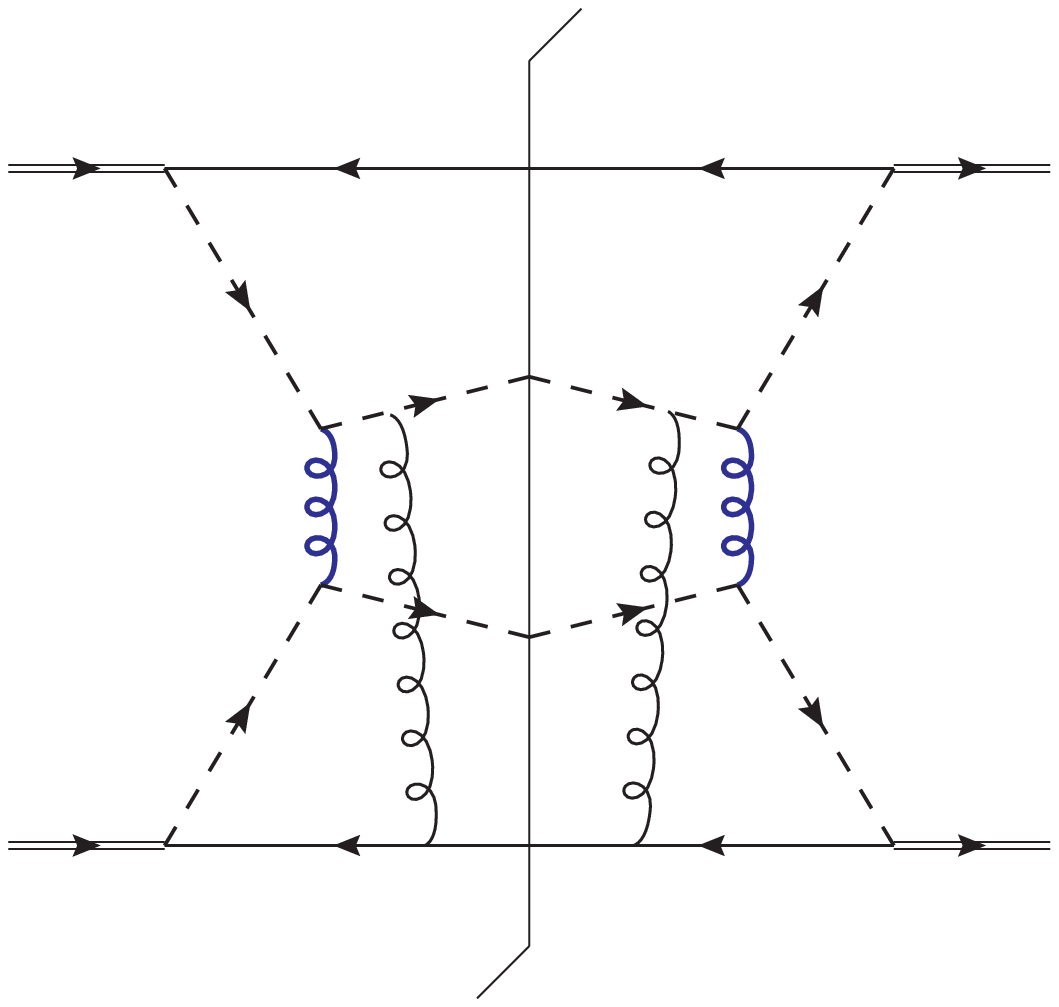}
  \\
    (a) & (b) & (c)
  \end{tabular}
  \caption{(Color online.)  Typical graphs for the exchange of two
    extra virtual gluons on opposite sides of the final-state cut,
    between the lower spectator and the active quark lines.  The
    classes of graph are: (a) Two gluons attaching to the outgoing
    quark of charge $g_1$.  (b) One gluon connecting to the $g_1$
    quark, one to one of the active $g_2$ lines.  (c) Both gluons
    connecting the the active $g_2$ lines.  The total number of graphs
    is 9.  }
  \label{fig:HHHH11}
\end{figure*}

\begin{figure*}
  \centering
  \begin{tabular}{c@{\hspace*{5mm}}c@{\hspace*{5mm}}c}
    \includegraphics[scale=0.4]{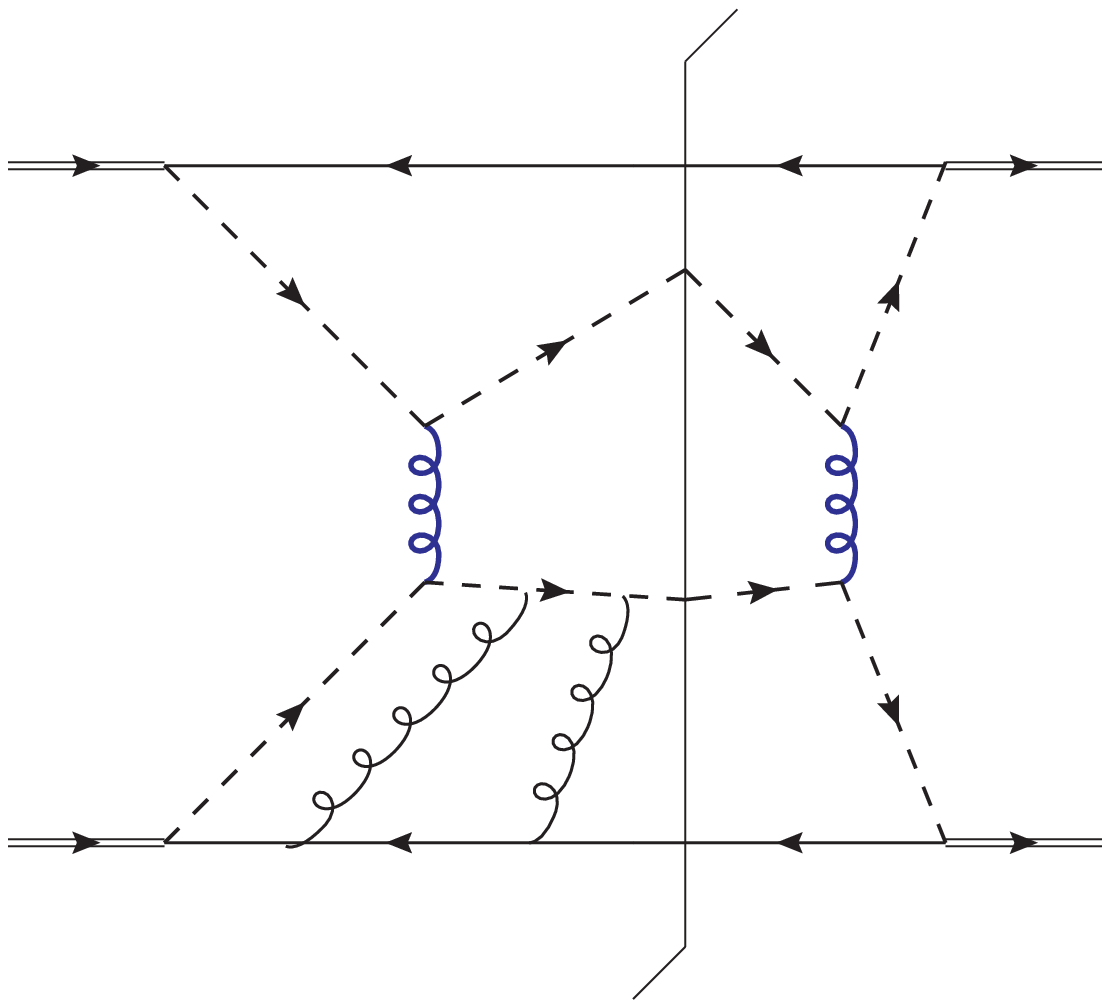}
  &
    \includegraphics[scale=0.4]{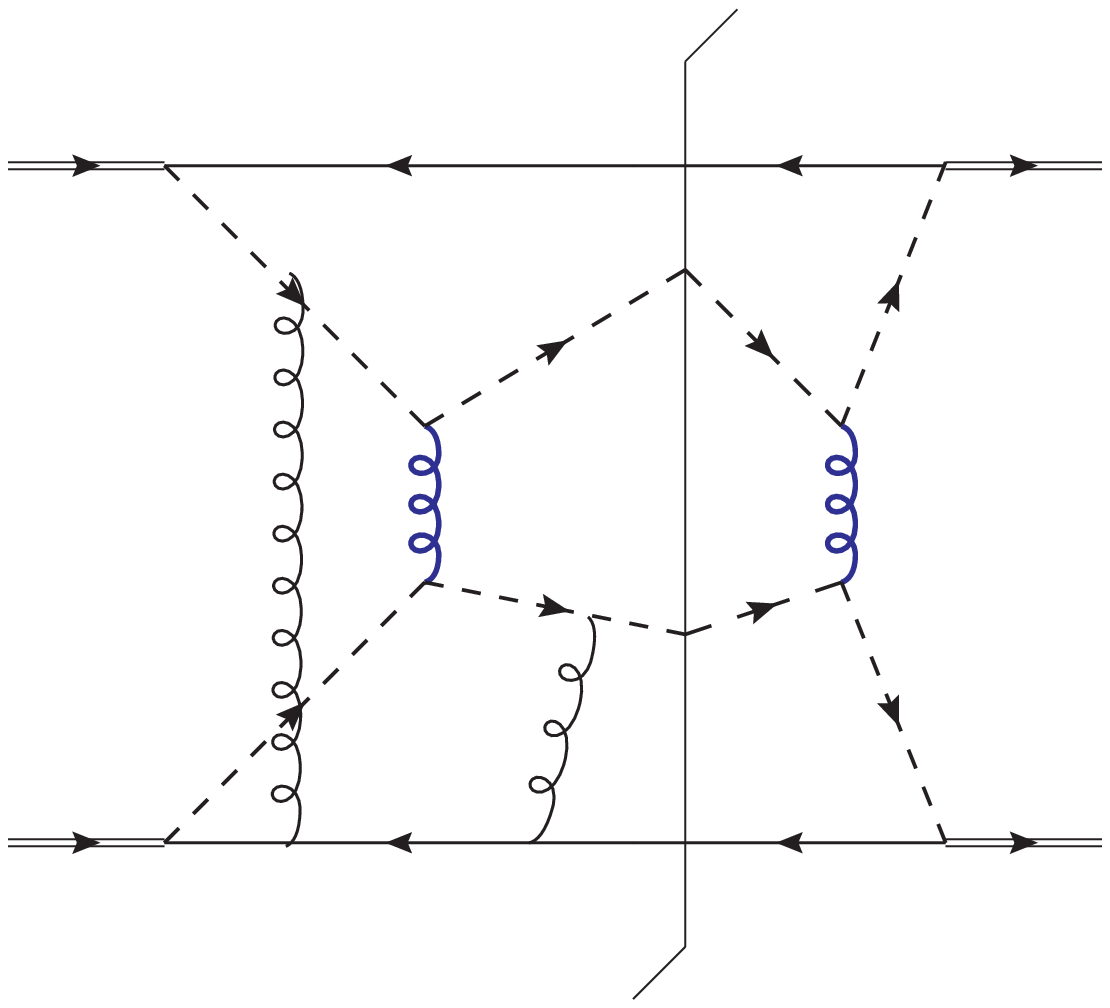}
  &
    \includegraphics[scale=0.4]{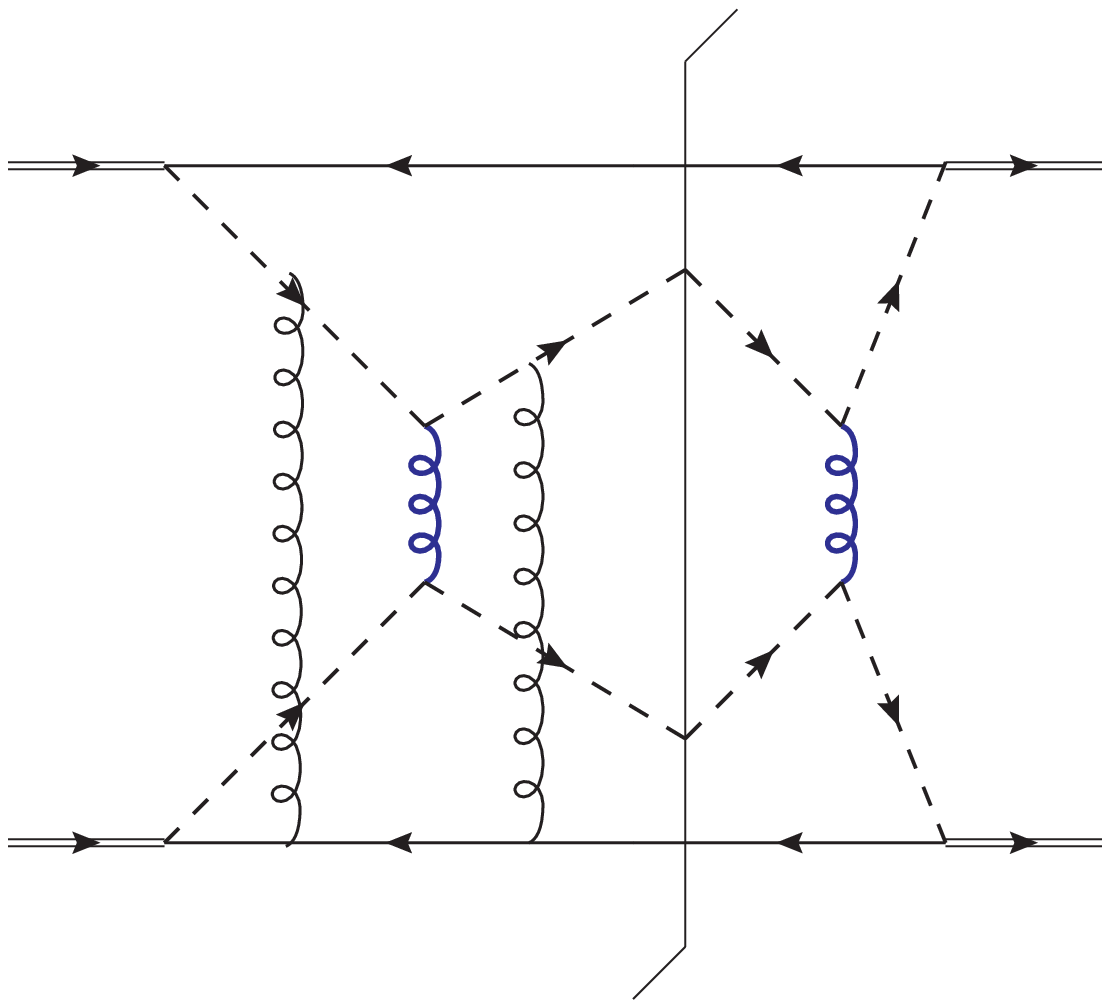}
  \\
    (a) & (b) & (c)
  \end{tabular}
  \caption{(Color online.)  Typical graphs for the exchange of two
    extra virtual gluons on one side of the final-state cut, between
    the lower spectator and the active quark lines.  The classes of
    graph are: (a) Two gluons attaching to the outgoing quark of
    charge $g_1$.  (b) One gluon connecting to the $g_1$ quark, one to
    one of the active $g_2$ lines.  (c) Both gluons connecting the the
    active $g_2$ lines.  The total number of graphs is 12, to which
    are to be added an equal number of Hermitian conjugate graphs.
  }
  \label{fig:HHHH2}
\end{figure*}

The lowest-order graph Fig.\ \ref{fig:HHHH0} is unambiguously
consistent with factorization, with the standard lowest-order value
for the hard scattering.  So we consider exchanges of one or more
extra gluons between the lower spectator line and the active partons
$k_2$, $k_3$, $k_4$, as in Figs.\ \ref{fig:HHHH1}, \ref{fig:HHHH11},
and \ref{fig:HHHH2}.  These graphs are among those including the gluon
exchanges whose sum must correspond to gluons attached to the Wilson
in the parton density for the lower hadron, if factorization is to
hold.  The graphs are obtained from Fig.\ \ref{fig:HHHH0} by inserting
virtual gluon lines between the lower spectator and the active
partons.  Note that graphs with gluons attached between the spectator
and the $k_1$ parton are unambiguously part of the parton density.

\subsection{Why just these graphs?}

Particularly with two extra gluons, Figs.\ \ref{fig:HHHH11} and
\ref{fig:HHHH2}, there are many more graphs than those we actually
examine.  A priori, it is conceivable that including other graphs
could change the results of the calculation, to be presented below,
and hence our conclusions as to factorization or nonfactorization.

It will be necessary justify our restriction to examining just the
graphs in Figs.\ \ref{fig:HHHH1}, \ref{fig:HHHH11}, and
\ref{fig:HHHH2}, together with certain related graphs.  The related
graphs for Fig.\ \ref{fig:HHHH1} are just Hermitian conjugates of
those shown.  For Figs.\ \ref{fig:HHHH11} and \ref{fig:HHHH2}, they
are those related by attaching the upper ends of the gluons in all
possible ways to the active partons of the specified charge.  In
addition, for Fig.\ \ref{fig:HHHH2}, there are the Hermitian conjugate
graphs.

The following general observations will assist in the justifications
of the choice of graphs.

As in \cite{Collins:2007nk}, we use light-front coordinates where the
lower hadron has large $+$ momentum and the upper hadron has large $-$
momentum.  The initial-state and final-state poles trap the $-$
component(s) of the momentum of the extra exchanged gluon(s) at small
values, but leave the possibility of deforming the $+$ momentum away
from the poles on the active parton lines.  Thus we can replace the
quark propagators that join the top ends of the gluon lines to the hard
scattering by eikonalized propagators, which, if factorization were to
hold, would correspond directly to the Feynman rules for Wilson lines.

The general form of factorization is that the differential cross
section is a convolution
\begin{equation}
\label{eq:factorization}
  d\sigma = H \otimes S \otimes C_1 \otimes C_2 \otimes C_3 \otimes C_4 + \mbox{power-suppressed},
\end{equation}
with a hard factor, a soft factor, and four collinear factors that
correspond to the observed particles with the same subscript numbers.
Standard power-counting results lead to a number of general results
for the regions that contribute to the leading power of $Q$, and how
they would give standard factorization, were it to be valid with the
known methods of proof.

Each region of a graph for the process gives a decomposition into
subgraphs whose momenta correspond to the factors in Eq.\
(\ref{eq:factorization}).  These subgraphs give contributions to the
factors, possibly after a summation by Ward identities, possibly after
some cancellations, and with the necessary subtractions to avoid
double counting.

The gluon is massive, and we have chosen the high-transverse-momentum
detected particles $k_3$ and $k_4$ to be almost back-to-back.  Then,
at leading power, the spectator lines are always collinear to their
parent hadron, and thus have low transverse momentum.  The active
partons indicated by the dashed lines are then always collinear in the
appropriate directions.  In the classes of graphs shown, the extra
exchanged gluons connect to the lower spectator line; these gluons can
only be either soft or collinear to one of the detected initial- or
final-state particles.

Here ``soft'' means soft the sense of \cite{collins_85_88}: all the
momentum components are much less than the hard scale $Q$.  Moreover,
under the conditions stated, including a nonzero gluon mass, soft
gluon lines connected to a spectator line are actually in the Glauber
region (as defined in \cite{Bodwin:1981fv,bodwin_85,collins_85_88}).
That is, the longitudinal components of gluon momentum are much less
than the transverse gluon momentum.

What makes the Glauber region central to issues of factorization
against nonfactorization is that the Ward identity arguments used to
obtain factorization do not apply in the Glauber region.  There are
two ways in which factorization can be nevertheless obtained.  One is
by a contour deformation out of the Glauber region, to a collinear
region or to a non-Glauber part of the soft region.  The other is by a
cancellation.  Note that real gluons are never Glauber.  Thus we
restrict our attention to virtual gluon contributions.

\subsection{Review: One extra gluon}

For one extra gluon \cite{Collins:2007nk}, we need the graphs of Fig.\
\ref{fig:HHHH1}.  Graphs with a virtual gluon connecting the upper
spectator line to the active partons are treated exactly similarly.
Graphs with a virtual gluon between the two spectators cancel after a
sum over cuts, as in factorization for the Drell-Yan process.  The
cancellation is between final states where the spectators have
different transverse momenta; thus it only occurs because the cross
section is fully inclusive in the two beam-fragmentation regions.  A
corresponding cancellation by sum over cuts does not apply to graphs
in Fig.\ \ref{fig:HHHH1}, because the relative transverse momentum of
the lines $k_3$ and $k_4$ is \emph{not} integrated over.

Up to an overall factor, the same for all three graphs, Fig.\
\ref{fig:HHHH1} gives
\begin{align}
\label{eq:eik1}
  E(l)
&=  \frac{g_2}{l^++i\epsilon}
  + \frac{g_2}{-l^++i\epsilon}
  + \frac{g_1}{-l^++i\epsilon}
\nonumber\\
&=
  -2\pi i g_2 \delta(l^+)
  + \frac{g_1}{-l^++i\epsilon}
\nonumber\\
&=
  -\pi i (2g_2+g_1) \delta(l^+)
  - \text{PV}\frac{g_1}{l^+}.
\end{align}
For the unpolarized cross section, only the real part is relevant,
i.e., the principal value in the last line.  It corresponds exactly to
the standard Wilson line, Fig.\ \ref{fig:pdf1}, with color charge
$g_1$.  

For the SSA we need only the imaginary part, from the delta-function
term.  Its coefficient is $2g_2+g_1$, whereas the standard Wilson line
in a standard parton density corresponds to a coefficient $g_1$, with
a sign depending on the direction of the Wilson line.  This is
incompatible with the structure required by standard factorization.

But the lowest order contribution to the SSA, from Fig.\
\ref{fig:HHHH0}, is zero.  So, if one were to \emph{ignore the
  possibility of exchanging even more gluons}, one might propose that
factorization works but with the hard scattering multiplied by a
process-dependent color factor.

\subsection{Two extra gluons on opposite sides of final-state cut}

Next, we examine the case of two extra virtual gluons, of momenta
$l_1$ and $l_2$.  A generalization of the arguments used with one
extra gluon shows that we can restrict our attention to graphs where
the gluons connect the lower spectator line with the $k_2$, $k_3$, and
$k_4$ external lines of the hard scattering, in all possible ways.  In
the case that the extra gluons are on \emph{opposite} sides of the
final-state cut, typical graphs are shown in Fig.\ \ref{fig:HHHH11}.
If factorization were valid, the sum of these graphs would correspond
to contributions to the parton density for hadron $H_1$ with two
gluons connecting the spectator line to the Wilson line.

We again apply the eikonal approximation to the attachments next to
the hard scattering, and the result is the integral over a common
factor, the same for all the graphs in Fig.\ \ref{fig:HHHH11}, and an
eikonal factor, which is a one-gluon eikonal --- as in Eq.\
(\ref{eq:eik1}) --- times a complex conjugate eikonal
\begin{align}
\label{eq:eik11}
   E(l_1)E(l_2)^*
= {}&
  g_1^2 \text{PV}\frac{1}{l_1^+l_2^+}
\nonumber\\&
  + i\pi g_1(2g_2+g_1) \delta(l_1^+) \text{PV}\frac{1}{l_2^+}
\nonumber\\&
  - i\pi g_1(2g_2+g_1) \text{PV}\frac{1}{l_1^+} \delta(l_2^+) 
\nonumber\\&
  + \pi^2 (2g_2+g_1)^2 \delta(l_1^+)\delta(l_2^+).
\end{align}

For the SSA we need the imaginary part of this product, the middle two
lines.  Since these are linear in the imaginary parts of the one-gluon
eikonal $E$, the imaginary part of the product gets the same color
enhancement factor $1+2g_2/g_1$ as in the exchange of one extra gluon.
Thus the SSA from these graphs is still consistent with the proposal
that factorization holds when the hard scattering is given the
color-enhancement factor.

\begin{figure*}
  \centering
  \begin{tabular}{c@{\hspace*{5mm}}c@{\hspace*{5mm}}c}
    \includegraphics[scale=0.4]{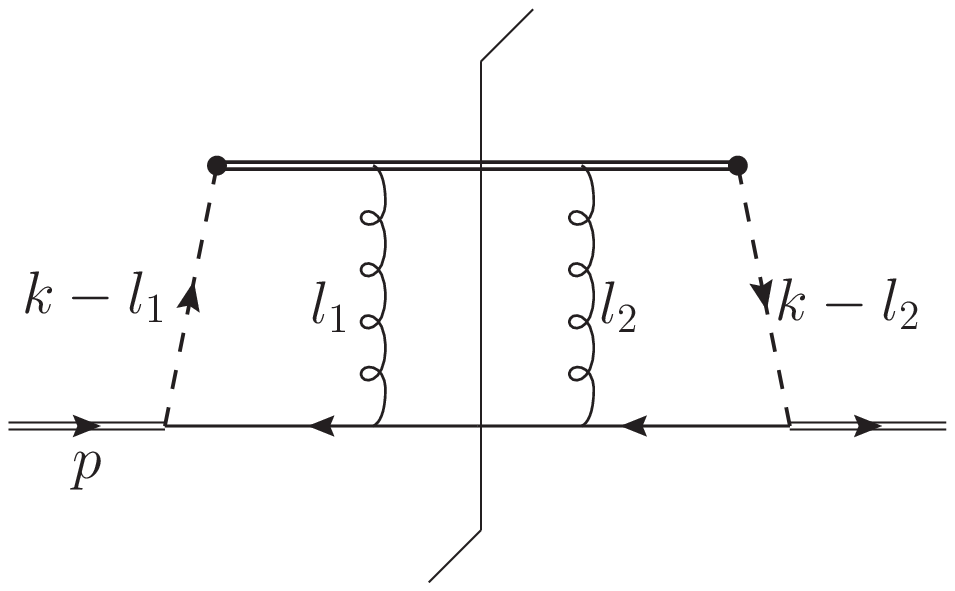}
  &
    \includegraphics[scale=0.4]{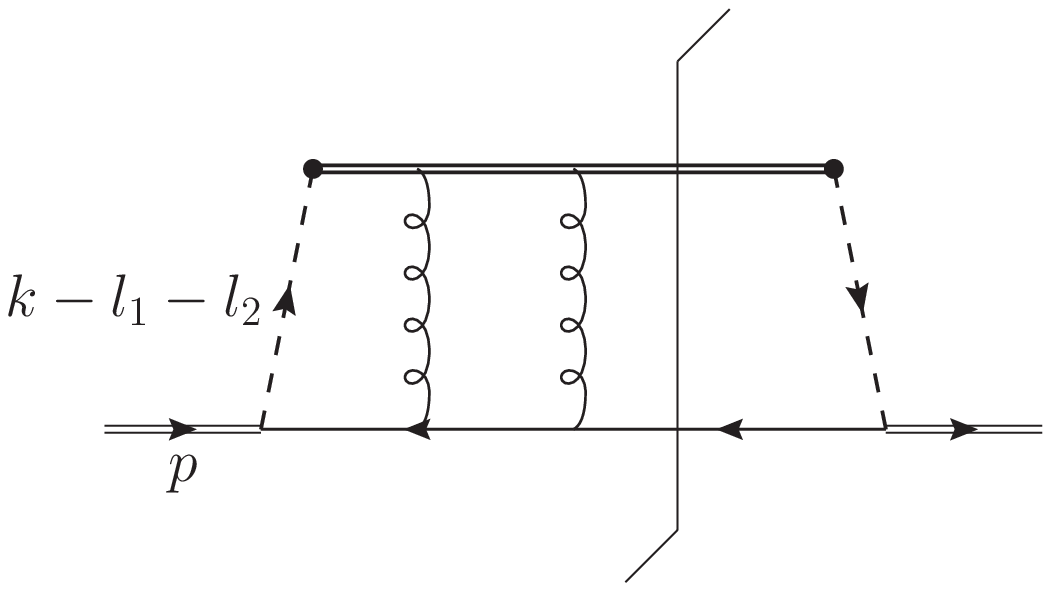}
  &
    \includegraphics[scale=0.4]{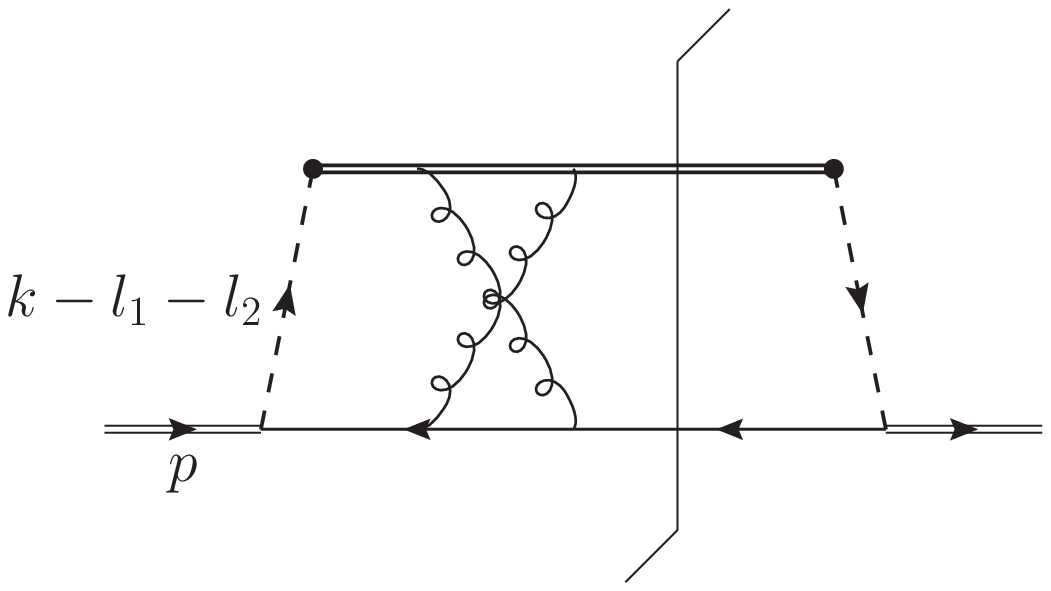}
  \\
    (a) & (b) & (c)
  \end{tabular}
  \caption{Virtual two-gluon-exchange corrections to parton density.
    Only graphs with gluons connecting the spectator line to the
    Wilson line are shown.  There are two further graphs of this type
    which are the Hermitian conjugates of graphs (b) and (c). }
  \label{fig:pdf2}
\end{figure*}

But this is not so for the unpolarized cross section, which comes from
the real part of (\ref{eq:eik11}), its first and last lines.  The
$g_1^2$ terms are, of course, just those that are consistent with
standard factorization, and correspond to the graph in Fig.\
\ref{fig:pdf2}(a) for the parton density.

The remaining terms provide what we can call the anomaly term
\begin{equation}
\label{eq:eik11.anom}
   \left. E(l_1)E(l_2)^* \right|_{\rm anom}
= 4\pi^2 g_2(g_2+g_1) \delta(l_1^+)\delta(l_2^+).
\end{equation}
This is evidently non-zero, and corresponds to a violation of
factorization for the unpolarized cross section.  

However, there are still more relevant graphs that give an anomaly,
those where the extra gluons are on the same side of the final-state
cut.  We will next analyze them, so that we can verify there is no
cancellation between the different sets of graphs.

\subsection{Two extra gluons on same side of final-state cut}

Fig.\ \ref{fig:HHHH2} illustrates the graphs with two extra virtual
gluons which are both on the same side of the final-state cut.  Each
graph is an integral over the product of a common factor and an
eikonal factor.  We now sum the graph-specific eikonal factors in
order to find the anomaly that is to be added to the graphs in Fig.\
\ref{fig:pdf2}(b) and (c) for the parton density.

When both gluons attach to the same line, there are two graphs, and in
an Abelian theory the eikonals combine in a simple way, e.g.,
\begin{widetext}
\begin{align}
  \frac{g^2}{ (-l_1^+-l_2^++i\epsilon)\, (-l_1^++i\epsilon) }  
  + \frac{g^2}{ (-l_1^+-l_2^++i\epsilon)\, (-l_2^++i\epsilon) }  
=
  \frac{g^2}{ (-l_1^++i\epsilon)\, (-l_2^++i\epsilon) } ,
\end{align}
where $g$ is $g_1$ or $g_2$.

The terms proportional to $g_1^2$ are exemplified in Fig.\
\ref{fig:HHHH2}(a), and they give exactly the result for DIS pdfs,
i.e.,
\begin{align}
   \frac{g_1^2}{ (-l_1^++i\epsilon)\, (-l_2^++i\epsilon) }  
=
  g_1^2 \text{PV} \frac{1}{l_1^+ l_2^+}  - \pi^2 g_1^2 \delta(l_1^+) \delta(l_2^+) 
  +i\pi g_1^2 \delta(l_1^+) \text{PV} \frac{1}{l_2^+}
  +i\pi g_1^2 \delta(l_2^+) \text{PV} \frac{1}{l_1^+}.   
\end{align}
The various terms proportional to $g_2^2$ give
\begin{align}
&  \frac{g_2^2}{ (l_1^++i\epsilon)\, (l_2^++i\epsilon) }  
  + \frac{g_2^2}{ (l_1^++i\epsilon)\, (-l_2^++i\epsilon) }  
  + \frac{g_2^2}{ (-l_1^++i\epsilon)\, (l_2^++i\epsilon) }  
  + \frac{g_2^2}{ (-l_1^++i\epsilon)\, (-l_2^++i\epsilon) }  
\nonumber\\
 &\hspace{1cm}
= 
  g_2^2 \left[ \frac{1}{l_1^++i\epsilon} + \frac{1}{-l_1^++i\epsilon}  \right]
\left[ \frac{1}{l_2^++i\epsilon} + \frac{1}{-l_2^++i\epsilon}  \right]
\nonumber\\
 &\hspace{1cm}
= 
   - g_2^2 (2\pi)^2 \delta(l_1^+) \delta(l_2^+).
\end{align}
This is nonzero, and contributes to the real part of the amplitude, so
it clearly gives a violation of factorization for the unpolarized
cross section.  It can be checked that the integrals over $l_1^-$ and
$l_2^-$ give a result that is real (and nonzero). Thus the real and
imaginary parts of the eikonal correctly indicate the real and
imaginary parts of the amplitude.
The terms proportional to $g_1g_2$ give
\begin{align}
&  \left[ \frac{g_2}{l_1^++i\epsilon} + \frac{g_2}{-l_1^++i\epsilon} \right]
  \frac{g_1}{-l_2^++i\epsilon}
 +  \left[ \frac{g_2}{l_2^++i\epsilon} + \frac{g_2}{-l_2^++i\epsilon} \right]
  \frac{g_1}{-l_1^++i\epsilon}
\nonumber\\
&\hspace{1cm}
=
  -2\pi i g_2 \delta(l_1^+) \frac{g_1}{-l_2^++i\epsilon}
  -2\pi i g_2 \delta(l_2^+) \frac{g_1}{-l_1^++i\epsilon}
\nonumber\\
&\hspace{1cm}
=
  - (2\pi)^2 g_1g_2 \delta(l_1^+) \delta(l_2^+) 
  + 2\pi i g_1g_2 \delta(l_1^+) \text{PV} \frac{1}{l_2^+}
  + 2\pi i g_1g_2 \delta(l_2^+) \text{PV} \frac{1}{l_1^+}.
\end{align}
\end{widetext}

Notice that the total of these is just the product of two one-gluon
eikonals $E(l_1)E(l_2)$.  This corresponds to exponentiation in a
Wilson-line operator, with the non-standard one-gluon term $E(l)$.
Thus we have verified factorization in the generalized sense of
\cite{bomhof_04,Bacchetta:2005rm,Bomhof:2006dp,Pijlman:2006tq}.  There
modified paths for the Wilson line operators are used instead of the
standard ones. 

As with the previous case, the imaginary part
\begin{align}
g_1^2(1+2g_2/g_1)  
\left[
  i\pi \delta(l_1^+) \text{PV} \frac{1}{l_2^+}
  + i\pi \delta(l_2^+) \text{PV} \frac{1}{l_1^+}
  \right]
\end{align}
is linear in the anomalous part of the one-gluon eikonal.  Thus it
continues to be the standard imaginary part times the lowest-order
color-enhancement factor for the SSA.  So \emph{at this order} we
still have consistency with the factorization proposed by Qiu,
Vogelsang and Yuan \cite{Qiu:2007ar,Qiu:2007ey} for the SSA.  However,
this is misleading as regards the status of factorization for the SSA.
Only with yet one more gluon will the iterated anomalous imaginary
part affect the SSA.

In contrast, for the unpolarized cross section, we have a high enough
order to get a real contribution from the iterated anomalous one-loop
eikonal.  It is easily checked that the anomalous term is the
negative of the anomalous term (\ref{eq:eik11.anom}) for the case that
the extra gluons are on opposite sides of the final-state cut.

\subsection{Total}

We have two sets of graphs, each of which evidently gives a nonzero
anomalous contribution, i.e., anomalous with respect to standard
factorization.  Our final step is to verify that there is no
cancellation.  We will also verify that we do get a cancellation if
the transverse momentum of the active parton $k_1$ is integrated over,
to correspond to the ordinary integrated parton density.  

Thus we have factorization violation when TMD densities are used, but,
at least at this order, we continue to have collinear factorization,
provided that we work with a cross section that is not sensitive to
partonic $k_T$. 

There are two common factors for every graph considered: these are the
hard scattering and the upper part of the graphs, which corresponds to
the parton density in the upper hadron, both the same as in the lowest
order graph, Fig.\ \ref{fig:HHHH0}.  The anomalous terms all
correspond to graphs for the lower parton density of the form of Fig.\
\ref{fig:pdf2}, but with the Wilson line factors replaced by the
relevant anomalous eikonal.  

The precise values of the graphs depend on the dynamics of the theory,
but to demonstrate that no cancellation follows from general
principles, it is sufficient to verify non-cancellation in a simple
model.  Since we are no longer concerned with an SSA, we choose a
simpler model than in \cite{Collins:2007nk}: We let all the lines that
model quarks and hadrons be scalars, with a hadron-quark-quark
coupling of $\lambda$.  We let $m_g$ and $m_q$ be the gluon and quark
masses, and we further simplify the kinematics by setting the hadron
mass $M$ to zero.

When the extra gluons are on opposite sides of the final-state cut, as
in Fig.\ \ref{fig:pdf2}(a), the anomalous term is
\begin{widetext}
\begin{align}
\label{eq:I1}
  I_1(k_T)
  ={}& 
  \frac{ \lambda^2g_1^2g_2(g_2+g_1) }{ (2\pi)^{12} }
  xp^+ \int dk^- \, d^4l_1 \, d^4l_2 \,
  \frac{ [2(p^+-k^+)+l_1^+] \, [2(p^+-k^+)+l_2^+] }
       { (l_1^2-m_g^2) \, (l_2^2-m_g^2) \, 
         [(k-l_1)-m_q^2+i\epsilon] \, [(k-l_2)-m_q^2+i\epsilon]
       }
\times \nonumber\\ & \times 
  \frac{ (2\pi)^2 \delta(l_1^+)\delta(l_2^+) ~
         2\pi\delta\big( (p-k)^2-m_q^2 \big)
       }
       { [(p-k+l_1)-m_q^2+i\epsilon] \, [(p-k+l_2)-m_q^2+i\epsilon] }
\nonumber\\ 
  ={}& 
  \frac{ \lambda^2g_1^2g_2(g_2+g_1) x(1-x) }{ 256\pi^7 }
  \int d^2l_{1T} \, d^2l_{2T} \, 
  \prod_{j=1,2} \frac{ k_T^2+m_q^2 }
               { (l_{jT}^2+m_g^2) \, [(k_T-l_{jT})^2+m_q^2] }.
\end{align}
The factor $xp^+$ in the first line is from the definition of a parton
density for a scalar quark.  When the extra gluons are on the same
side of the final-state cut, Fig.\ \ref{fig:pdf2}(b), (c), and their
Hermitian conjugates, the anomalous term is similarly
\begin{equation}
\label{eq:I2}
  I_2(k_T)
  =
  \frac{ -\lambda^2g_1^2g_2(g_2+g_1) x(1-x) }{ 256\pi^7 }
  \int d^2l_{1T} \, d^2l_{2T} \, 
  \frac{1}
       { (l_{1T}^2+m_g^2) \, (l_{2T}^2+m_g^2) \,
         [(k_T-l_{1T}-l_{2T})^2+m_q^2] \, (k_T^2+m_q^2)  
       }.
\end{equation}
\end{widetext}
Integrating $I_1(k_T)+I_2(k_T)$ over all $k_T$ gives zero.  Thus,
certainly at this order, the factorization anomaly cancels in
quantities that are not sensitive to partonic $k_T$ and so can use
integrated parton densities.  

To verify that the cancellation is not point-by-point in $k_T$, we
simply verify that $I_1(k_T)+I_2(k_T)$ is nonzero for one value of
$k_T$.  For example, with a certain amount of effort, it can be proved
analytically that $I_1(0)+I_2(0) < 0$.

\section{Conclusions}

We have calculated explicitly that the graphs with two extra exchanged
gluons coupling the spectator to the hard scattering give a result
inconsistent with standard $k_T$-factorization for the unpolarized
cross section.  This is caused by the imaginary parts of the eikonal
propagators for the partons at the hard scattering.  The mismatch with
standard factorization occurs because there are both initial- and
final-state active partons in the process considered, production of
hadrons in hadron-hadron collisions.

Any cancellation with other graphs for the parton density would
require cancellations within the integral over single graphs, or
between graphs for the parton density of very different topology (and
hence with very different dependence on the kinematic variables).  We
have verified that the cancellation does not happen in a specific
case.

For the imaginary part, appropriate to the SSA, the iterated anomalous
imaginary part of the one-gluon eikonal does not contribute at the
order of perturbation theory that we examined.  So at this order there
is no \emph{explicit} contradiction with the proposal of
\cite{Qiu:2007ar,Qiu:2007ey}, where the hard scattering is modified by
a color factor; such a contradiction would need a yet higher order in
gluon exchange.  But it must be emphasized that the general conversion
of extra gluon exchanges to the Wilson line form uses standard Ward
identities.  These are of a form that does \emph{not} give the
modified hard scattering.  Therefore, as explained in
\cite{Collins:2007nk}, the extra color factor at one-gluon order is by
itself sufficient to show that the conversion of extra gluon exchanges
to standard Wilson lines fails.

Since the regions of momentum investigated are appropriate to
nonperturbative physics, they correspond to a strong effective
coupling in QCD.  Thus the fact that nonfactorization occurs two
orders of perturbation theory beyond the lowest order for a process is
not indicative of any special suppression.

We have verified that at least within our example a cancellation of
the anomalous term does occur if partonic $k_T$ is integrated over.
This verifies that collinear factorization continues to be valid.
However, resummation methods that handle the back-to-back region are
endangered, as are any other methods that are sensitive to the
detailed transverse structure of the final state.

There will presumably be some further non-factorization effects when
spectator-spectator interactions are included.  Although such
interactions cancel in the Drell-Yan process
\cite{bodwin_85,collins_85_88}, the results here show that the
conditions for the cancellation may no longer occur when cross
sections to hadrons are examined.

\begin{acknowledgments}
  This work was supported in part by the U.S.\ Department of Energy
  under grant number DE-FG02-90ER-40577.  I would like to thank the
  following for useful conversations: A.  Bacchetta, J.  Qiu, T.
  Rogers, A. Stasto, G. Sterman, M. Strikman, W. Vogelsang, F.  Yuan,
  and the participants of the workshop ``Transverse momentum, spin,
  and position distributions of partons in hadrons'' at the European
  Centre for Theoretical Studies in Nuclear Physics and Related Areas
  in Trento, Italy, June 2007.
  
  The figures in this paper were made using JaxoDraw \cite{jaxodraw}.
\end{acknowledgments}


\end{document}